\documentclass[floatfix,aps,prl,twocolumn,groupedaddress]{revtex4}
\usepackage{graphicx}
\begin{document}

%\preprint{}

\title{Active Plasmonics: Controlling Signals in Au/Ga Waveguide using Nanoscale Structural Transformations}


\author{A.V. Krasavin}
\email[]{avk@soton.ac.uk}
\author{N.I. Zheludev}
%\thanks{}
\affiliation{School of Physics and Astronomy, University of
Southampton, SO17 1BJ, UK}


\date{\today}

\begin{abstract}
We develop a new concept for active plasmonics exploiting
nanoscale structural transformations which is supported by
rigorous numerical analysis. We show that surface
plasmon-polariton signals in a metal-on-dielectric waveguide
containing a gallium section a few microns long can be effectively
controlled by switching the structural phase of gallium. The
switching may be achieved by either changing the waveguide
temperature or by external optical excitation. The signal
modulation depth could exceed 80 percent and switching times are
expected to be in the picosecond-microsecond time scale.
\end{abstract}

\pacs{Here should be pacs}
%\keywords{}

\maketitle

%\section{}
%\subsection{}
%\subsubsection{}

We are entering the age of integrated photonic devices for signal
and information processing when planar waveguides and photonic
crystal structures are being intensively investigated as primary
solutions for guiding light in such devices. There may however, be
another means of making highly integrated optical devices, with
structural elements smaller than the wavelength, enabling strong
guidance and manipulation of light using metallic and
metallodielectric nanostructures. Here plasmon-polariton waves,
i.e optical excitations coupled with collective electronic
excitations are used as information carriers. A range of very
promising nanostructures that guide plasmon-polariton waves
\cite{R1, R2, Z8} are currently being investigated. Surface
plasmon polaritons in gold films can propagate for tens of microns
and may be guided by structuring the metal film and creating
polaritonic band gap materials \cite{Z10}. Propagating
plasmon-polariton excitations in nanostructured metal films are
therefore clearly emerging as a new information carrier for
highly-integrated photonic devices \cite{R5}. However, we will
only be able to speak about 'plasmonics' in the same way that we
speak about 'photonics' when efficient techniques for active
manipulation of surface polariton-polariton sygnals are
identified.

Here we propose a concept of active nanoscale functional elements
operating with surface plasmon-polariton signals. Our approach
takes advantage of the most characteristic features of surface
plasmon-polaritons, namely their localization in nanometer thick
surface layers of metal, and the fact that their propagation
depends strongly on the metal's electronic properties. These
features are exploited by combining them with the recently
developed concept of achieving nanoscale photonic functionality
using structural phase transitions in polyvalent metals, an idea
which has already been shown to offer all optical switching at
milliwatt power levels in thin films \cite{Z15} and nanoparticles
\cite{Z16}, and promises a new type of photodetector \cite{Z17}.
Here we show that surface plasmon-polariton signals in metallic on
dielectric waveguides containing a gallium section a few microns
long can be effectively controlled by switching the crystalline
gallium section of the waveguide from one structural phase to
another. The switching may be achieved by both changing the
waveguide temperature and by external optical excitation.
\begin{figure*}
    \includegraphics[width=140mm]{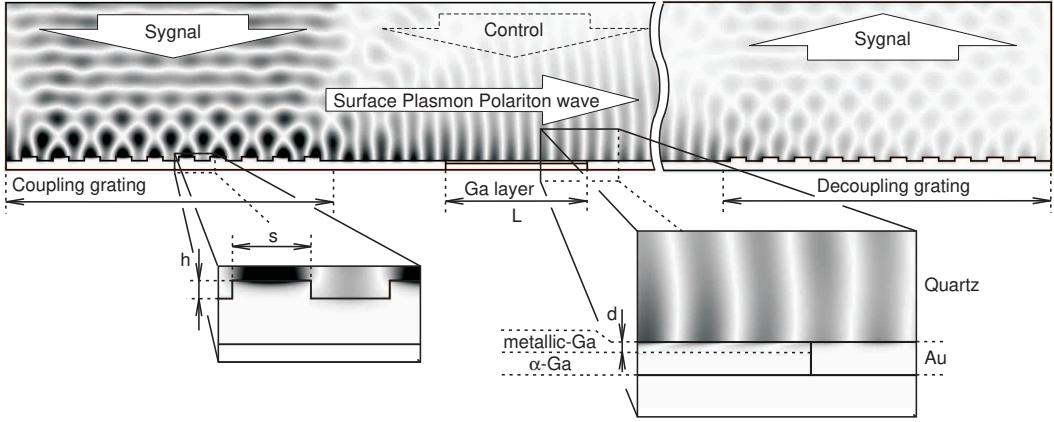}\\
  \caption{A SPP gold-on-quartz waveguide containing a gallium switching section. The metallic film is at the bottom of the quartz substrate. Field
mapping shows the magnitude of the magnetic filed.}\label{Fig1}
\end{figure*}
Metallic Gallium is a uniquely suitable material for this
application. It is known for its polymorphism \cite{R9}. In
$\alpha$-gallium, the stable 'ground-state' phase \cite{Ziii}
molecular and metallic properties coexist - some inter-atomic
bonds are strong covalent bonds, forming well-defined Ga$_2$
dimers (molecules), and the rest are metallic bonds. The structure
is highly anisotropic, with much better thermal and electrical
conductivity in the 'metallic planes' than along the covalent
bonds. Remarkably, $\alpha$-gallium has a very low melting point,
$29.8^\circ$C. The covalent bonding leads to a strong optical
absorption peak centered at 2.3 eV and spreading from
approximately 0.68 eV ($\sim$310 nm) to the mid-infrared part of
the spectrum. Optical properties of the $\alpha$-Ga and more
metallic phases, metastable at normal conditions, are greatly
different, in terms of the dielectric coefficients of liquid
gallium and solid gallium they are huge, at a wavelength of
1.55$\mu$m we have
$|\varepsilon_{liquid}-\varepsilon_\alpha|\sim180$. The metallic
metastable phase (quasi-melt) may be achieved by simple heating,
or by light absorption through a non-thermal "optical melting"
mechanism via destabilization of the optically excited covalent
bonding structure \cite{Z15}. Whatever the mechanism of phase
transition is it is a surface mediated effect and develops as
propagation of the new metallic phase from the surface into the
bulk of the semiconductor-like $\alpha$-phase. As the phase
transition only involves a few tens of atomic layers of gallium at
the interface it is highly reproducible and fully reversible and
could run for millions of cycles without noticeable changes. High
quality gallium interfaces with silica may be achieved using
various techniques, from squeezing molten gallium to ultra-fast
pulsed laser deposition \cite{Zvii}.

To evaluate the potential switching characteristics of the Surface
Plasmon Polariton (SPP) waveguide we numerically modelled it using
the Finite Elements Method. We investigated a gold film waveguide
containing gallium section of length L on a quartz substrate. To
model coupling and decoupling of optical radiation to and from the
waveguide two ten element meander gratings were placed at both
ends of the structure (Fig.\ref{Fig1}). Although optimization of
coupling and decoupling efficiency was not the prime objective of
this study, we found that efficiencies in excesses of 20$\%$ for
coupling and 80$\%$ for decoupling could be achieved by the
meander gratings with $s/h=1/5$, while coupling levels above
30$\%$ are possible with gratings of complex profile. In such a
waveguide the SPP wave propagates at the interface between the
metal film and silica substrate, through the gold and gallium
sections. SPP decay length in a continuous gold/silica waveguide
is $53\mu$ for the excitation wavelength of $1.31\mu$. In the
waveguide containing Ga section, the transmitted SPP wave
attenuates due to the mismatch of dielectric characteristics at
the Au-Ga boarder and losses which are much higher in gallium than
in gold. We modelled that the gallium section of the waveguide may
be converted from the ground $\alpha$-phase to the metallic
liquid-like phase. We also modelled that this change could take
place as interface metallization: a thin layer of metallic gallium
of thickness $d$ develops at the silica interface (see
Fig.\ref{Fig1}). As $\alpha$-gallium is a highly anisotropic
material (space group $Cmca$), we performed calculations of the
waveguide transmission (disregarding coupling-decoupling losses)
for all main crystalline orientations of the Ga film at the
interface in the range of optical excitation wavelengthes from
0.9$\mu$ to 2.0$\mu$.
\begin{figure}[h]
    \includegraphics[angle=270,width=85mm]{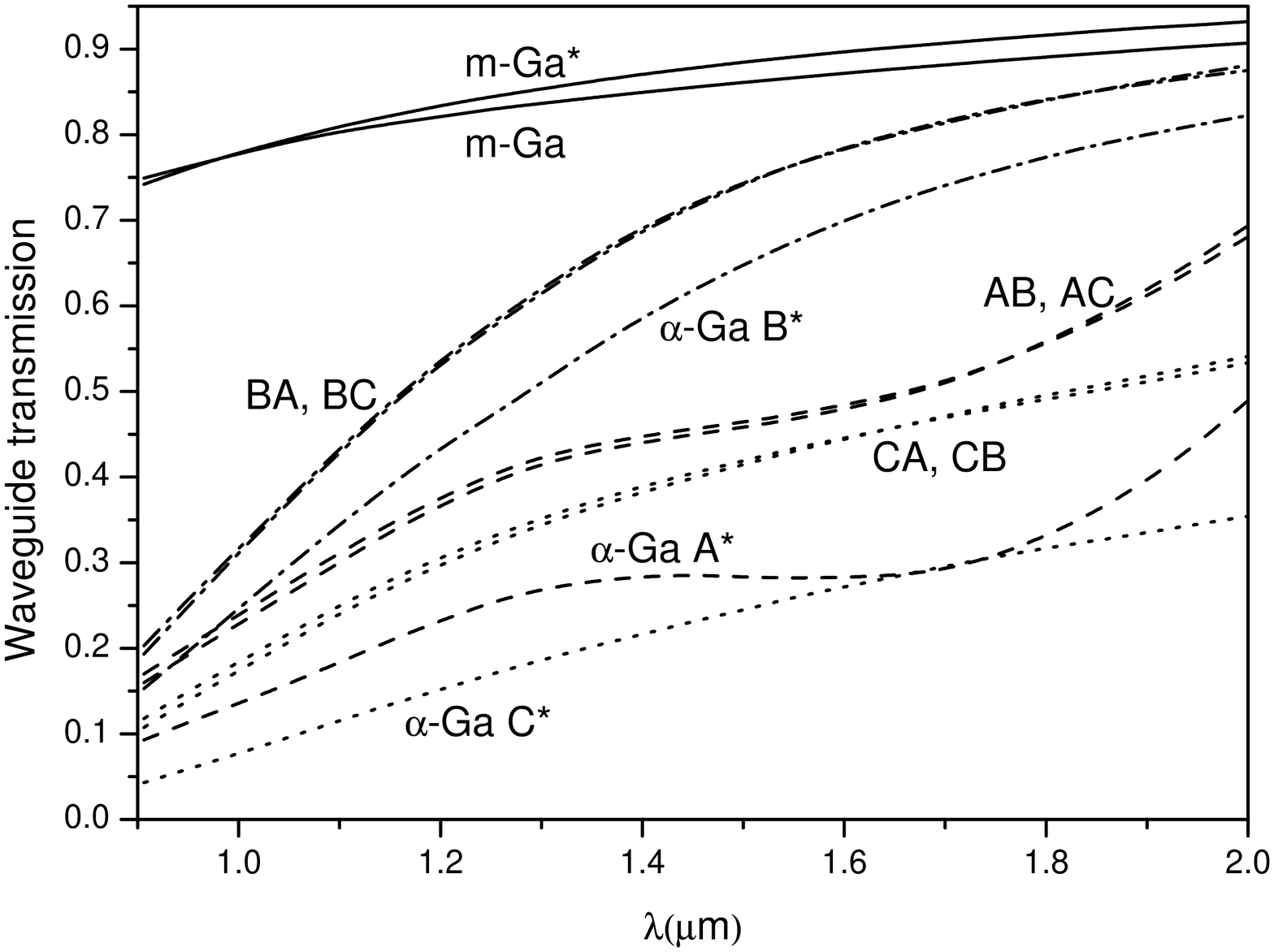}\\
  \caption{Waveguide transmission as a function of wavelength
  for different phases and crystalline orientations of gallium film.
  Curves without asterisks are calculated in FEM numerical simulation,
  while curves marked with asterisks are obtained analytically.
  Compare the waveguide transmissions for the gallium section
  in the metallic phase ($m-Ga$ and $m-Ga^*$, solid lines)
  and crystalline phases (dashed lines).}\label{Fig2}
\end{figure}
Fig.~\ref{Fig2} shows the results of these calculations at
different wavelengths of incident light for a waveguide containing
gallium section of length L = $2.5\mu$, assuming that the gallium
section is a homogenously crystal of given orientations or is a
fully molten isotropic liquid phase. The graph also show the
transmission levels calculated using analytical theory which
accounts for absorption in an isotropic infinite homogeneous
waveguide \cite{Z23} for three main values of gallium's
crystalline dielectric coefficient and for the dielectric
properties of the liquid state. The following notation was used:
curve AB corresponds to gallium crystalline structure with its
A-axis laying in the interface plane perpendicular to the
direction of the SPP propagation and its B-axis perpendicular to
the strip. Similar notation applies to curves BA/BC and CA/CB. The
values of the complex dielectric coefficients of gallium and gold
were taken from ref.~\cite{A1} and \cite{A2}. Although the results
of numerical analysis and analytical calculations show similar
spectral trends, the magnitudes of transmission coefficients are
somewhat different for these two approaches. We argue that the
discrepancy reflects limitations of the analytical theory which
ignores the reflection of SPPs at the gold/gallium boarder. More
importantly, however, the analytical theory also ignores the
"hopping" of the electromagnetic wave across the narrow gallium
strip followed by re-coupling of the excitation into the SPP on
the other side of the strip. These effects are very pronounced in
numerical mapping of the fields and change the waveguide
transmission in comparison with predictions of the analytical
theory.
\begin{figure}[h]
    \includegraphics[angle=270,width=85mm]{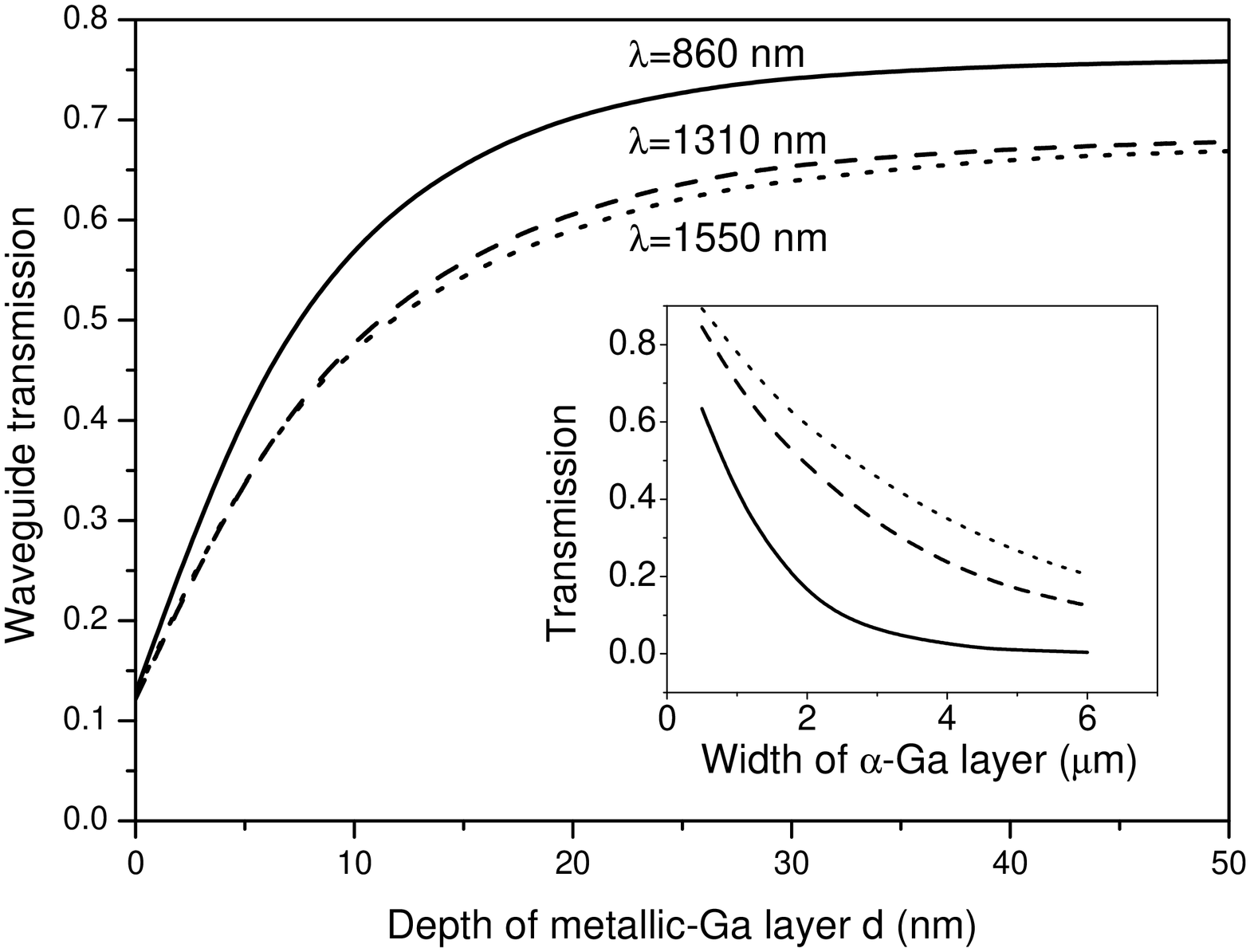}\\
  \caption{Waveguide transmission as a function of the depth $d$ of the metallic gallium layer.
  Insert shows the dependence of transmission on the width $L$ of gallium section.}\label{Fig3}
\end{figure}
Therefore our calculations show that the waveguide transmission
strongly depends on the structural phase of the gallium section
and changing of the phase could be used for active controlling of
the transmission. As the structural transformation in gallium is a
surface mediated effect, two phases of gallium may co-exist at the
interface, with a thin layer of metallic phase sandwiched between
silica and $\alpha$-gallium. In fact, melting of gallium takes the
form of "surface melting" with the molten phase thickness steadily
increasing with temperature in a narrow temperature corridor just
below the gallium bulk melting point. Similarly, light-induced
metallization also develops as a surface mediated effect and
starts from the interface while the equilibrium thickness of the
metallized layer $d$ may be controlled by the light intensity.
This gives a possibility of continuous, "analog" control of the
waveguide transmission. To analyze this effect we calculated the
waveguide transmission for different thicknesses of metastable
gallium phase up to $d=$60nm for a number of incident wavelengths
(see Fig.\ref{Fig3}). For illustrative purposes the gallium strip
was taken to be polycrystalline $\alpha$-gallium: isotropic with
dielectric constants averaged over crystal directions. In can be
seen that the presence of the metallic liquid Ga layer in the
section of only a few microns long and depth of just $d$=30 nm can
dramatically change the transmission of plasmons through the
waveguide.

One therefore can envisage an active plasmonic devise in which SPP
transmission is controlled by the waveguide temperature in the
range of a few degrees or external optical stimulation. In the
near-infrared part of the spectrum a typical fluence of optical
excitation needed for converting of $\alpha$-Ga phase to a
metallic phase of several tens of nanometers deep is about 10
mJ/cm$^2$ \cite{Z15}. For a section of gallium waveguide 2.5$\mu$m
x 2.5$\mu$m the optical energy required for high-contrast
switching would be in the order of 10pJ. The envisaged application
of the control light signal to the waveguide is schematically
presented on Fig.1 by a dashed arrow. When the excitation is
terminated the molten/metallic layer rapidly recrystallizes into
the ground $\alpha$-phase. The intrinsic switch-on time was
measured for a gallium-quartz interface, and was found to be 2-4
ps \cite{Zvii} and we expect that the SPP switch-on time will also
be in the scale of a few ps. We anticipate the SPP switch-off time
to be in the microsecond-nanosecond time scales \cite{Zvi}.
Notably, this is 4-8 orders of magnitude faster than the currently
achieved response time of opto-mechanical switching microdevices.

The authors  acknowledge the support of the Science and
Engineering Research Council (U.K.) and fruitful discussion with
A.Zayats.

\end{document}